\documentstyle[prc,twocolumn,aps,epsf]{revtex}
\begin{document}
\twocolumn[\hsize\textwidth\columnwidth\hsize\csname
@twocolumnfalse\endcsname

\draft
\title{Role of spectroscopic factors in the potential-model
description of the $\bbox{^7}$Be$\bbox{(p,\gamma)}$$\bbox{^8}$B
reaction}
\author{Attila Cs\'ot\'o}
\address{Department of Atomic Physics, E\"otv\"os University,
P\'azm\'any P\'eter s\'et\'any 1/A, H--1117 Budapest, Hungary}
\date{August 30, 1999}

\maketitle

\begin{abstract}
In standard potential-model descriptions of the 
$^7{\rm Be}(p,\gamma){^8{\rm B}}$ reaction the $^7{\rm Be}+p$
spectroscopic factors ${\cal S}$ appear in the cross section. We
argue that the microscopic substructure effects which are
represented by ${\cal S}$ are short-ranged and cannot affect the
asymptotic normalization of the wave function. We believe that the 
standard way of describing reactions in a potential model may be 
incorrect and the low-energy cross section should not depend on  
${\cal S}$ in the case of external capture reactions, like 
$^7{\rm Be}(p,\gamma){^8{\rm B}}$
\end{abstract}
\ \\
]

\narrowtext

Recently the $^7{\rm Be}(p,\gamma){^8{\rm B}}$ radiative capture
reaction has been studied extensively both experimentally and
theoretically. This interest is rooted in the fact that the 
$^8$B produced by this reaction in our sun is the main source of
the high-energy solar neutrinos \cite{Bahcall}. The high-energy 
solar neutrino flux is directly proportional to the
low-energy ($E_{\rm cm}=20$ keV) astrophysical cross section factor,
$S_{17}(E)$, of $^7{\rm Be}(p,\gamma){^8{\rm B}}$. Among the
recent experimental results are a new direct measurement of the
low-energy cross section by using a $^7$Be target
\cite{Hammache}, the determination of $S_{17}(E)$ from the
inverse process, the Coulomb dissociation of $^8$B \cite{CD},
and the utilization of transfer reactions in order to determine
the asymptotic normalization constant of the bound-state $^8$B
wave function \cite{Liu,Azhari}, which in turn can be used to 
extract $S_{17}(0)$. On the theoretical side, the capture
reaction has been studied recently in $^7{\rm Be}+p$ potential 
models \cite{Alex,Typel}, in three-body models
\cite{Grigorenko}, in shell-models \cite{Bennaceur}, and in 
microscopic cluster models \cite{clust,Desc}. Interesting 
results came also from the R-matrix study of the experimental
data \cite{Barker95}, from the investigation of the
energy-dependence of $S(E)$ \cite{Jennings}, and from the
studies of the asymptotic normalization constants of the $^8$B 
wave function \cite{Xu,Timof}.
Yet, despite all these and other advances, $S_{17}(0)$ is still
the most uncertain input parameter in solar models
\cite{Adelberger}. 

In the present work we would like to clarify a few points of
this problem in connection with the potential models.

At solar energies the $^7{\rm Be}(p,\gamma){^8{\rm B}}$ reaction
takes place deep below the Coulomb barrier. This means that the
capture cross section receives contributions almost exclusively
from those parts of the initial scattering and final bound state
wave functions that describe large $^7{\rm Be}-p$ separations.
At low energies the scattering wave functions are almost fully
known, as the phase shifts practically coincide with the hard
core phase shifts. The asymptotic behavior of the bound state
wave function is also known, as it is proportional to the
Coulomb-Whittaker function,
\begin{equation}
\chi^{\rm bound}_I(r)=\bar c_I W^+_{-\eta,l+1}(kr)/r, \ \ \
r\rightarrow \infty,
\label{anc}
\end{equation}
where $\eta$ is the Coulomb parameter, $l$ is the relative
angular momentum between $^7$Be and $p$, and $I=1, 2$ is the 
channel spin, which comes from the coupling of the $^7$Be spin 
and the spin of the proton. Therefore, the zero-energy 
$^7{\rm Be}(p,\gamma){^8{\rm B}}$ cross section depends only on
the asymptotic normalization constants $\bar c_I$ \cite{Mukh,Xu}. 
Using a generic formula, which is specified later for the various 
models, this means 
\begin{equation}
S_{17}(0)=N(\bar c_1^2+\bar c_2^2) \ \ {\rm eVb}.
\label{gen}
\end{equation}
(We note, that the different notations followed in various papers are 
slightly confusing. Our $\bar c$ quantity is the equivalent of $\beta$ in 
Ref.\ \cite{Xu}, while the spectroscopic factor ${\cal S}$, see 
below, is denoted by ${\cal J}$ there.)\,  
We would like to emphasize that we use the Eq.\ (\ref{anc}) 
definition of the asymptotic normalization constant in all 
cases. In the case of a potential-model description, $\chi$ 
is the so-called single-particle wave function, while in 
the case of a microscopic model, $\chi$ is the wave function 
describing the relative motion between $^7$Be and $p$.

The precise value of $N$ depends on the details of the
scattering wave function. In the case of our scattering state  
coming from the microscopic cluster model \cite{clust}, which
corresponds roughly to $r_c=2.4$ fm hard-sphere radius  
\cite{Jennings}, $N$ is 37.8, therefore 
\begin{equation}
S_{17}(0)^{\rm micr.}=37.8(\bar c_1^2+\bar c_2^2) \ \ {\rm eVb}.
\label{mic}
\end{equation}
Note that in Refs.\ \cite{clust} the integration of the cross 
section was not done to a sufficiently large radius. All $S_{17}(0)$ 
values given there should be increased by roughly 0.4 eVb. 
Note also that in Ref.\ \cite{Alex} the hard-core scattering states,
used in a potential model, were chosen to be in sync with those
coming from Refs.\ \cite{clust}, but once again the integration 
distance was too short. Thus, the corrected  
$S_{17}(20\,{\rm keV})=37.2(\bar c_1^2+\bar c_2^2)$ eVb relation
found there, is in agreement with Eq.\ (\ref{mic}). 

\begin{figure}
\centerline{\epsfxsize 13cm \epsfbox{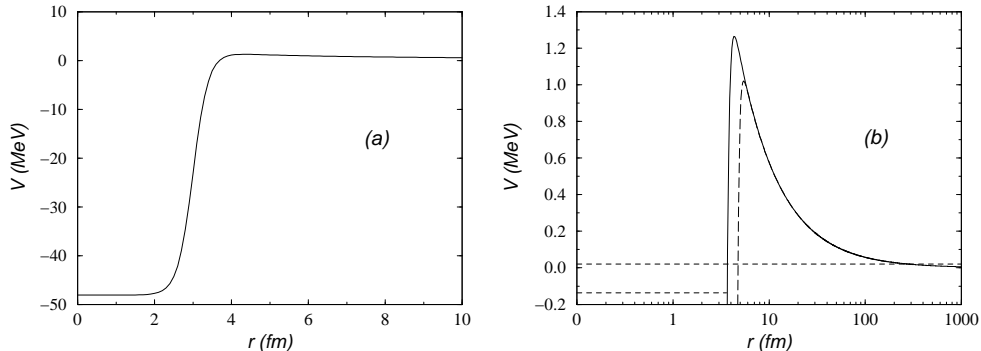}}
\caption{Schematic local potential between $^7$Be and $p$ (a), and 
its barrier region magnified (b). The horizontal dashed lines 
denote the energy of a 20 keV proton incident on $^7$Be and the 
binding energy, $-137$ keV, of $^8$B, relative to the $^7{\rm Be}+p$ 
threshold, respectively. The long-dashed line shows the Coulomb barrier 
of a potential with a somewhat larger radius.} 
\label{fig1}
\end{figure}

The peripheral nature of the $^7{\rm Be}(p,\gamma){^8{\rm B}}$
reaction is illustrated in Fig.\ \ref{fig1}. A schematic local
potential is shown between $^7$Be and $p$. One can see that at
20 keV, which is the most effective reaction energy in
our sun, the proton hits the Coulomb barrier at about 250 fm. It
has to tunnel through a huge barrier in order to allow the
capture to take place. Therefore, the cross section is really
sensitive almost exclusively to the asymptotic parts of the wave
functions. One can also see in Fig.\ \ref{fig1} that the
asymptotic normalization of the bound state wave function is
most sensitive to the radius of the $^7{\rm Be}-p$ potential. A
slightly bigger radius (shown by the long-dashed line in Fig.\
\ref{fig1}; for the sake of illustration the change in the
radius is strongly exaggerated) leads to a smaller and narrower
barrier, and thus to a significantly larger tunneling
probability, which gives a larger cross section \cite{clust,tunnel}. 

In Eqs.\ (\ref{gen}-\ref{mic}) there is one point which is not yet
specified, namely the full normalization of the bound-state wave
function. In the case of an 8-body model of the reaction, the
full 8-body wave function should be normalized to unity.
However, the $^7{\rm Be}-p$ relative motion wave functions, as 
one-dimensional functions which contain the $\bar c$ constants,
are obviously not normalized the same way (see below). In the
case of a potential model, the effects of the internal structure
of $^7$Be, which are neglected in the model, has to be taken into
account in some implicit way. Conventionally this is done
through the spectroscopic factors ${\cal S}$. The
spectroscopic amplitude functions, $g$, of the $^7{\rm Be}+p$
configuration in $^8$B are given \cite{Varga} as
\begin{equation}
g({\bf r})=\langle \Psi_{\bf r}\vert \Psi^{\rm {^8B}}\rangle ,
\label{spa}
\end{equation}
where $\Psi^{\rm {^8B}}$ is the normalized antisymmetrized 
8-body wave function of $^8$B , while $\Psi_{\bf r}$ is defined 
as 
\begin{equation}
\Psi_{\bf r}= {\cal A}\Big [\Phi^{^7{\rm Be}}\Phi^p 
\delta({\bf r}-\bbox{\rho })\Big ].
\label{psir}
\end{equation}
Here ${\cal A}$ is the intercluster antisymmetrizer between
$^7$Be and $p$, $\Phi^{^7{\rm Be}}$ and $\Phi^p$ are the
normalized antisymmetrized internal wave function of $^7$Be and
a spin-isospin eigenstate of the proton, respectively, 
and $\bbox{\rho}$ is the relative coordinate between $^7$Be and
$p$. The quantum numbers carried by $g$, like the channel spin
$I$, the angular momentum coupling, etc., are not indicated here
for simplicity. The spectroscopic factor is given as 
\begin{equation}
{\cal S}=\int\vert g({\bf r})\vert ^2 d{\bf r}.
\label{spic}
\end{equation}
This quantity is a measure of the cluster substructure effects 
which are neglected in a potential model, and can be calculated
using a microscopic model, like the shell model or the cluster  
model, or can be extracted from nuclear reaction measurements.

The various quantities that are calculated from potential
models, like the decay widths of resonances, cross sections,
etc., contain ${\cal S}$ in order to take into account the
effects of the neglected microscopic substructure. In other
words, the norm of the potential-model wave function is assumed
to be different from unity, depending on how large the neglected
microscopic effects are. In potential models, the    
$^7{\rm Be}(p,\gamma){^8{\rm B}}$ cross section contains ${\cal
S}$ and thus the Eqs.\ (\ref{gen}-\ref{mic}) expressions are 
modified as 
\begin{equation}
S_{17}(0)^{\rm pot.}=37.8(\bar c_1^2{\cal S}_1+\bar c_2^2{\cal S}_2) 
\ \ {\rm eVb}
\label{pot}
\end{equation}
(here the same hard-core scattering state is assumed as in the
microscopic model),
where ${\cal S}_1$ and ${\cal S}_2$ are the 
channel-spin spectroscopic factors. It is important to note that
this way of taking into account the microscopic
effects relies on the assumption that these effects can be 
handled separately from the calculation of the matrix element 
of the cross section. We note also, that most of the 
potential-model calculations generate the $I=1$ and $I=2$ 
channel wave functions of the $^8$B ground state separately, 
not in a correct coupled-channel description. In any case, 
the total single-particle wave function (containing both the 
$I=1$ and $I=2$ components) is assumed here to be normalized 
to unity.  

\begin{figure}
\centerline{\centerline{\epsfxsize 7.5cm \epsfbox{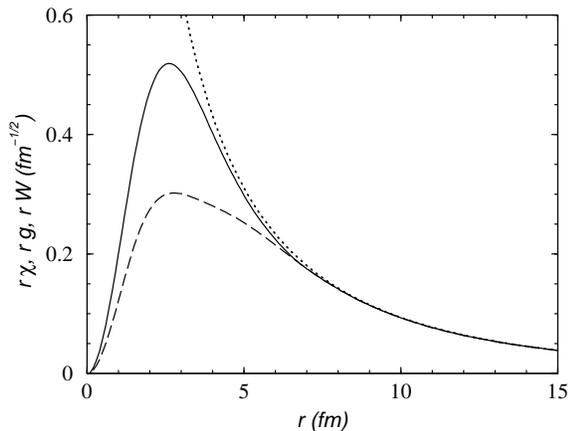}}}
\caption{The reduced $^7{\rm Be}+p$ relative motion wave function 
($r\chi$, dashed line) and spectroscopic amplitude ($rg$, solid line)  
of $^8$B in the $I=2$ channel, coming from the 
microscopic cluster model of Refs.\ \protect\cite{clust}. The dotted 
line is the corresponding $^7{\rm Be}+p$ Coulomb-Whittaker function 
multiplied by the asymptotic normalization constant, $\bar c_2=0.763$.}
\label{fig2}
\end{figure}

We would like to argue, however, that we believe that the
conventional definition of the cross section in the potential
model might be incorrect. The effects of the microscopic
substructure, which leads to the appearance of the spectroscopic 
factors in the cross section formula, are short-distance
corrections because they originate from short-range effects, 
like the antisymmetrization.
Therefore, these effects should only affect the internal parts
of the wave functions and should not modify the asymptotic
normalization constants. To illustrate this, in Fig.\ \ref{fig2}
we show the $^7{\rm Be}-p$ relative-motion wave function $\chi$ 
of $^8$B in the $I=2$ channel, coming from the microscopic
cluster model of Refs.\ \cite{clust} (dashed line). The
MN effective nucleon-nucleon interaction was used and the total
wave function contained only $({^4{\rm He}}+{^3{\rm He}})+p$
terms with cluster size parameters $\beta=0.4$ fm$^{-2}$. Note
that this relative-motion wave function is the one behind the
antisymmetrizer in the $^8$B wave function, thus its norm, which
is not one, has no physical meaning. Also shown in Fig.\
\ref{fig2} are the spectroscopic amplitude $g(r)$ (solid line)
defined in Eq.\ (\ref{spa}), and the Coulomb-Whittaker function
of the $^7{\rm Be}-p$ relative motion multiplied by the
asymptotic normalization constant $\bar c_2=0.763$ (for $I=1$, 
$\bar c_1=0.302$), coming from the model (dotted line). 
The spectroscopic amplitude was calculated using the procedure 
discussed in Ref.\ \cite{li6}. 
As one can see, the relative motion function $\chi$ and the
spectroscopic amplitude $g$ coincides beyond $r\approx 7$ fm.
The difference between the two functions in the internal region
gives a measure of the antisymmetrization effect. 

The effect
of taking into account the microscopic substructure in the
potential model would be similar (although much smaller, because
the norm of $g$ is close to one) on the potential-model wave 
function. Therefore, it seems to us that the usual way of
treating microscopic effects in the potential model, through the
spectroscopic factors, cannot be right. Multiplying the
potential-model wave function by $\sqrt{\cal S}$ modifies it not
only in the internal region but asymptotically as well. We
suggest that the correct way to take into account the effects of
the microscopic substructure in the potential model would be to
use the spectroscopic amplitude functions in the expressions of
the various matrix elements. In certain cases where the internal
parts of the wave functions play the major role, like the decay
widths of resonances or the cross sections of non-peripheral
reactions, the results would be close to those coming from the
conventional definition, because of the Eq.\ (\ref{spic})
relation. However, in certain cases, like the present
peripheral reaction cross section, where only the asymptotic
parts of the wave functions are important, there would be no
effect coming from the difference between the wave
function and $g$. If our suggestion turns out to be correct,
then the Eq.\ (\ref{pot}) cross section formula for potential
models should not contain the spectroscopic factors. 

We realize of course that
our present arguments are rather heuristic. A thorough study of
the connection between microscopic and macroscopic approaches to
capture reactions, similar to that presented in Ref.\
\cite{Varga} for nuclear structure, would be highly welcome. 

Using the spectroscopic amplitudes, one of which is shown in
Fig.\ \ref{fig2}, we calculated the spectroscopic factors
predicted by our present cluster model. They are ${\cal
S}^J_{I_7,I}={\cal S}^2_{3/2,2}=0.915$ and ${\cal S}^2_{3/2,1}= 
0.205$. Here $I_7$ and $I$ are the spin of $^7$Be and the
channel spin, respectively, while $J$ is the total spin of the 
$^8$B ground state. The spectroscopic factor of the state which
contains the excited state of $^7$Be ($I_7=1/2$) is 
${\cal S}^2_{1/2,1}=0.25$ in our model. The total spectroscopic
factor corresponding to $I_7=3/2$ and $J=2$ is ${\cal S}=
{\cal S}^2_{3/2,2}+{\cal S}^2_{3/2,1}=1.12$, in good agreement
with the shell-model predictions of Ref.\ \cite{Alex}. We note
that the spectroscopic factors given in Refs.\
\cite{Barker95,Barker} erroneously do not take into account
the $A/(A-1)=8/7$ center-of-mass correction factor
\cite{Dieperink}. The corrected numbers are ${\cal S}=1.177$,
1.166, and 1.143 for the CK, B, and K shell-model interactions,
again in good agreement with Ref. \cite{Alex}. 

We calculated the spectroscopic factors also for the other
models and interactions used in Refs.\ \cite{clust} and found
them to be between 1.07 and 1.12. One can observe a correlation
between $S_{17}(0)$ and ${\cal S}$ if the peak position of $g$
is roughly the same: a larger ${\cal S}$ leads to a slight
increase of $g$ in the asymptotic region, and thus to a bigger
$S_{17}(0)$. All our results with the MN and MHN interactions
fit into this trend. In the case of the V2 interaction the peak
position of $g$ is shifted to a slightly larger radius, which
makes the corresponding ${\cal S}=1.07-1.1$ small, compared to
the large $S_{17}(0)=28.8-29.8$ eVb cross section, predicted 
by the V2 force. We note that our
spectroscopic factors are slightly smaller than, but compatible
with the shell-model result, ${\cal S}=1.1-1.2$ \cite{Alex}. 
A conservative estimate of ${\cal S}$, based on the shell model 
and the cluster model, could be around $1.05-1.2$. 

In summary, we presented circumstantial evidences which
indicate that the conventional way of treating microscopic
substructure effects in potential models may be wrong. We
believe that the correct way of handling those effects would be
the use of the spectroscopic amplitude functions, instead of the
potential-model wave functions and spectroscopic factors. This
would lead to a zero-energy cross section for peripheral
reactions, which is independent of the spectroscopic
factors. 

\acknowledgments

This work was supported by OTKA Grants F019701, F033044, 
and D32513, and by the Bolyai Fellowship of the Hungarian 
Academy of Sciences. Some preliminary investigations related to
the present work were benefited from discussions with R.~G.
Lovas and K. Varga.

\vskip 1.5cm

\mediumtext

\vskip 1.0cm

\narrowtext


\begin{references}
\bibitem{Bahcall} J.~N. Bahcall, {\it Neutrino
astrophysics} (Cambridge University Press, Cambridge,
1989).
\bibitem{Hammache} F. Hammache {\it et al.}, Phys. Rev.
Lett. {\bf 80}, 928 (1998).
\bibitem{CD} Motobayashi {\it et al.}, Phys. Rev.
Lett. {\bf 73}, 2680 (1994); N. Iwasa {\it et al.}, Journ.
Phys. Soc. Japan {\bf 65}, 1256 (1996); T. Kikuchi {\it et
al.}, Eur. J. Phys. {\bf 3}, 213 (1998).
\bibitem{Liu} W. Liu {\it et al.}, Phys. Rev. Lett. {\bf 77},
611 (1996).
\bibitem{Azhari} A. Azhari {\it et al.}, Phys. Rev. Lett. {\bf
82}, 3960 (1999).
\bibitem{Alex} B.~A. Brown, A. Cs\'ot\'o, and R. Sherr, Nucl.
Phys. {\bf A597}, 66 (1996).
\bibitem{Typel} H. Esbensen and G.~F. Bertsch, Nucl. Phys.
{\bf A600}, 37 (1996); S. Typel, H.~H. Wolter and G. Baur, Nucl. 
Phys. {\bf A613}, 147 (1997).
\bibitem{Grigorenko} L.~V. Grigorenko, B.~V. Danilin, V.~D.
Efros, N.~B. Shul'gina, and M.~V. Zhukov, Phys. Rev. C {\bf 57},
R2099 (1998); Phys. Rev. C, in press.
\bibitem{Bennaceur} K. Bennaceur, F. Nowacki, J. Okolowicz, 
and M. Ploszajczak, J. Phys. G {\bf 24}, 1631 (1998); 
Nucl. Phys. {\bf A651}, 289 (1999).
\bibitem{clust} A. Cs\'ot\'o, K. Langanke, S.~E. Koonin, and
T.~D. Shoppa, Phys. Rev. C {\bf 52}, 1130 (1995); A. Cs\'ot\'o, 
Phys. Lett. B {\bf 394}, 247 (1997); A. Cs\'ot\'o and 
K. Langanke, Nucl. Phys. {\bf A636}, 240 (1998).
\bibitem{Desc} P. Descouvemont and D. Baye, Phys. Rev. C 
{\bf 60}, 015803 (1999).
\bibitem{Barker95} F.~C. Barker, Nucl. Phys. {\bf A588}, 693 
(1995).
\bibitem{Jennings} B.~K. Jennings, S. Karataglidis, 
and T.~D. Shoppa, Phys. Rev. C {\bf 58}, 3711 (1998).
\bibitem{Xu} H.~M. Xu, C.~A. Gagliargi, R.~E. Tribble, A.~M. 
Mukhamedzhnov, and N.~K. Timofeyuk, Phys. Rev. Lett. {\bf 73}, 
2027 (1994).
\bibitem{Timof} N.~K. Timofeyuk, D. Baye, and P.
Descouvemont, Nucl. Phys. {\bf A620}, 29 (1997); N.~K. 
Timofeyuk, Nucl. Phys. {\bf A632}, 19 (1998).
\bibitem{Adelberger}E.~G. Adelberger {\it et al.}, Rev. Mod. 
Phys. {\bf 70}, 1265 (1998).
\bibitem{Mukh} A.~M. Mukhamedzhanov and N.~K. Timofeyuk, JETP 
Lett. {\bf 51}, 282 (1990); Sov. J. Nucl. Phys. {\bf 51}, 431 
(1990).
\bibitem{tunnel} A. Cs\'ot\'o, Heavy Ion Phys. {\bf 6}, 103 
(1997) [nucl-th/9704053]; nucl-th/9712033.
\bibitem{Varga} K. Varga and R.~G. Lovas, Phys. Rev. C {\bf 43},
1201 (1991).
\bibitem{li6} A. Cs\'ot\'o and R.~G. Lovas, Phys. Rev. C {\bf 46}, 
576 (1992).
\bibitem{Barker} F.~C. Barker, Aust. J. Phys. {\bf 33}, 177 
(1980).
\bibitem{Dieperink} A.~E.~L. Dieperink and T. de Forest, Phys.
Rev. C {\bf 10}, 543 (1974).
\end{references}
\end{document}